\documentclass[table]{ccjnl}
\usepackage{lipsum,amsmath}
\usepackage{cuted}
\usepackage{bm}
\usepackage{bbm}
\usepackage{graphicx}
\usepackage{array}
\usepackage{makecell}
\graphicspath{{figures/}}
\usepackage{color}

\title{Prediction-based Hybrid Slicing Framework for Service Level Agreement Guarantee in Mobility Scenarios: A Deep Learning Approach}
\author{Heng Zhang\inst{1,2}, Guangjin Pan\inst{1,2}, Shugong Xu\inst{1,2}, Shunqing Zhang \inst{1,2}, Zhiyuan Jiang\inst{1,2},\corinfo{shugong@shu.edu.cn}}
\receiveddate{Jan. 14, 2022}
\reviseddate{xxx. xx, 2022}
\Editor{xxx xxx}

\address[1]{School of Communication \& Information Engineering, Shanghai University, Shanghai 200444, China}
\address[2]{Shanghai Institute for Advanced Communication and Data Science (Shanghai University), Shanghai 200444, China}

\begin{document}
\maketitle

\begin{abstract}
Network slicing is a critical driver for guaranteeing the diverse service level agreements (SLA) in 5G and future networks. Inter-slice radio resource allocation (IS-RRA) in the radio access network (RAN) is very important. However, user mobility brings new challenges for optimal IS-RRA. This paper first proposes a soft and hard hybrid slicing framework where a common slice is introduced to realize a trade-off between isolation and spectrum efficiency (SE). To address the challenges posed by user mobility, we propose a two-step deep learning-based algorithm: joint long short-term memory (LSTM)-based network state prediction and deep Q network (DQN)-based slicing strategy. In the proposal, LSTM networks are employed to predict traffic demand and the location of each user in a slicing window level. Moreover, channel gain is mapped by location and a radio map. Then, the predicted channel gain and traffic demand are input to the DQN to output the precise slicing adjustment. Finally, experiment results confirm the effectiveness of our proposed slicing framework: the slices’ SLA can be guaranteed well, and the proposed algorithm can achieve the near-optimal performance in terms of the SLA satisfaction ratio, isolation degree and SE.
\end{abstract}
\section{Introduction}
\label{Introduction}

With the emergence of 5G telecommunication technology, cellular networks are envisioned to cater services to a wide variety of innovative vertical applications, such as Cellular Vehicle-to-Everything (C-V2X), augmented/virtual reality (AR/VR), with heterogeneous performance requirements including high data rates, ultra-low latency and high reliability \cite{foukas2017network}. Such highly diverse performance requirements of these new services impose a challenge for 5G in terms of scalability, availability, and cost-efficiency \cite{Peter2017_CM}.

Network slicing is recognized as a promising technique to guarantee differentiated service quality of service (QoS) and service level agreements (SLAs). Since it can enable multiple logical networks corresponding to different network services run on top of a common physical network infrastructure such that the slices can be customized to satisfy various SLAs through virtualization, isolation techniques \cite{8685766}. From a perspective of radio resource management, the fundamental challenge of network slicing lies in the trade-off of isolation and resource efficiency. On the one hand, to achieve non-interference between slices, the slicing system intends to ensure complete isolation between network slices. On the other hand, inherent radio spectrum scarcity promotes that all slices share a limited radio resource on-demand to ensure efficient utilization. Therefore, inter-slice radio resource allocation (IS-RRA) in the radio access network (RAN) becomes an open technical challenge \cite{ISRRA2020WC}.

\subsection{Related Work}
{The development of core network slicing is more mature and can be implemented through advanced computing, caching, and virtualization container technologies\cite{Peter2017_CM}. In contrast, there are still challenges in the design of radio management of RAN slicing that need to be studied. As mentioned before, these challenges include the utilization efficiency of radio resources, the performance isolation between slices, as well as the dynamics of service traffic flow\cite{ISRRA2020WC}. Recently, deep reinforcement learning (DRL), a promising artificial intelligence tool, has been widely applied for network slicing \cite{rli2018access,jie2021tcom,TVT2019LIANG,DRL-eMBBURLLC-TWC2021,DRLLSTM,sun2019mix,myWCL2021,liu2020edgeslice,rongpeng2021tvt,chen2019DRL,JSAC2020GANDRL} to address the above challenges due to its capacity to solve model-free problems.}

\cite{rli2018access} investigated the application of DRL in solving radio resource slicing under dynamic traffic demands, and try to maximize the users' QoS and spectrum efficiency (SE). The results exhibited the advantage of DRL in solving model-free resource allocation problems. Moreover, the authors in \cite{jie2021tcom} proposed {a DRL-driven hierarchical control strategy to guarantee the long-term QoS and SE.} Similarly, \cite{TVT2019LIANG} proposed a collaborative learning framework consisting of supervised learning in conjunction with DRL to perform large and small time-scale resource allocation, respectively. In \cite{DRL-eMBBURLLC-TWC2021}, the authors proposed a model-based DRL algorithm where eMBB resource allocation and URLLC preemptive scheduling were jointly optimized without considering the performance isolation between slices. {To address the dynamic nature of the environment, the authors of \cite{DRLLSTM} incorporated the long short-term memory (LSTM) into DRL to track user mobility.} Furthermore, \cite{sun2019mix} and \cite{myWCL2021} developed DRL methods to heterogeneous networks scenarios to solve joint user association and network slicing {to achieve network-level QoS guarantees and maximize SE.} Similarly, the authors of \cite{liu2020edgeslice} applied DRL in the alternating direction method of multipliers (ADMM) model to realize a decentralized radio, transport and computing resource orchestration of slices. {Innovatively,} \cite{rongpeng2021tvt} combined the advantages of DRL and graph attention networks to solve the joint handover and slicing problems in a dense cellular network scenario. 

Furthermore, {some works take into account the issues faced by the practical deployment of DRL, including convergence speed and performance loss in the training stage.} Based on \cite{rli2018access}, \cite{chen2019DRL} proposed a faster convergence DRL scheme by integrating discrete normalized advantage functions (DNAF) and the deterministic policy gradient descent (DPGD) algorithm. {To reduce the bad effects of randomly exploration action}, \cite{JSAC2020GANDRL} proposed a generative adversarial network-powered deep distributional Q network (GAN-DDQN) to train DQN offline. 

{Based on our literature review and analysis, the above works on network slicing still have the following limitations.
\begin{itemize}
    \item {In existing works, the fixed assignment slicing scheme can achieve perfect performance isolation among different slices. However, it is prone to result in low resource efficiency. On the other hand, the shared-based slicing scheme can maximize resource utilization, but requires complex algorithms to ensure the slices' SLA, and cannot guarantee the performance isolation. Therefore, \textit{how can we realize a trade-off between resource efficiency and slices' isolation in a general way is still an open question.}}
     
    \item{To improve the policy in the unknown environment, a DRL algorithm will try some actions randomly to estimate the reward of different actions. During explorations, the DRL algorithm may try some bad actions, which will deteriorate the QoS significantly and may lead to unexpected accidents in real systems. Thus, \textit{how guaranteeing the network performance during the exploration stage is a bottleneck for applying DRL in practical systems.}}
     \item{In a high dynamic scenario, the mobility of users could exacerbate service request volatility and make the pre-allocated radio resource of each slice inefficient. However, most of the existing works do not consider the high mobility scenario. Therefore, \textit{it is critical for the network to intelligently adjust resource slicing to provide seamless services in a high mobility environment.}}
\end{itemize}}
\subsection{Contributions}
In this paper, we focus on an OFDMA-based cellular network with user mobility. To address the above challenges, we first propose a soft and hard hybrid slicing framework to find a trade-off between isolation and resource efficiency and facilitate slices' SLA guarantee in the exploration phase. Meanwhile, we design a customized deep learning framework consisting of the LSTM and DQN. Specially, LSTM networks are used to perform network state-awareness at a large timescale, and DQN is adopted to achieve the precise adjustment of resources based on the predicted network state. The major contribution of the paper are summarized as follows:
\begin{itemize}
    \item Proposing a soft and hard hybrid slicing framework. It divides the resource into two parts, i.e., the hard and soft parts corresponding to dedicated resources for each slice and shared resources for all slices. On the one hand, the common slice setting significantly reduces the SLA violation in the DRL exploration phase. On the other hand, with reasonable resource allocation, the hybrid slicing framework can maximize resource efficiency while guaranteeing SLA under a required isolation degree constraint.
    \item Design of a joint LSTM-based network state prediction and DQN-based slicing control strategy. In the proposed two-step {deep} learning (DL)-based solution, LSTM networks are utilized to predict each user's traffic demand and location. Moreover, channel gain is mapped by location and a radio map. Then, the predicted future channel gain and traffic demand are input to the DQN to give an accurate slicing adjustment to adapt {to} user mobility. 
    \item Performance evaluation of the proposed DL-based hybrid framework. To give an accurate evaluation, we give three representative comparison algorithms: the \textbf{Optimal} algorithm, the state-of-the-art algorithms \textbf{Hard-LSTM-A2C} \cite{DRLLSTM} and {\textbf{Hard-DQN} \cite{sun2019mix}}, whose details are explained in Section \ref{sec:experi}. The numerical results illustrate that the proposed method outperforms Hard-DQN and Hard-LSTM-A2C, and achieves near-optimal performance.

\end{itemize}

The rest of the paper is organized as follows. Section \ref{SYSMODEL} {introduces} the communication, slice and SLA model. Section \ref{sec:proformu} first proposes the soft and hard hybrid slicing framework and then based on { the} hybrid slicing framework, IS-RRA is formulated as a utility maximization problem. Section \ref{solution} proposes a two-step DL-based algorithm, which combines LSTM-based network state prediction and DQN-based resource allocation of slices. The numerical results
are given in Section \ref{sec:experi}, followed by the conclusion
in Section \ref{sect:conc}.

\section{System Model}
\label{SYSMODEL}
\begin{figure*}[t]
    \centering
    \includegraphics[width=0.85\textwidth]{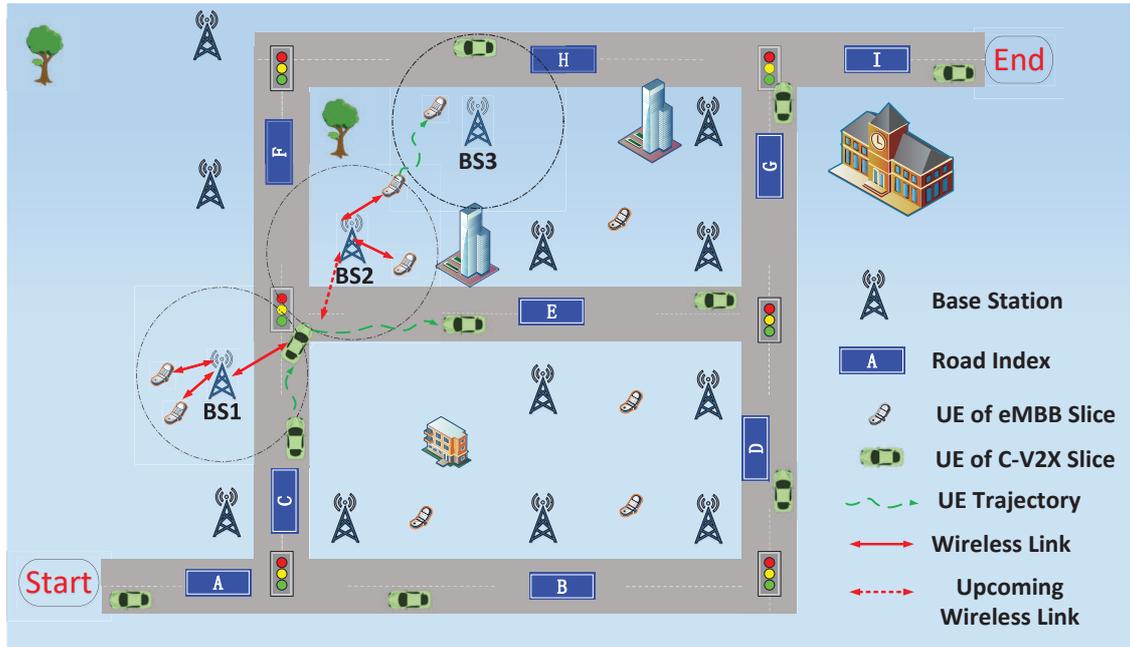}
    \caption{Overview of the  system scenario.}
    \label{fig:sys}
\end{figure*}
\subsection{Communication Model}

\begin{table}[t]
\centering
\caption{Description of key notations.}
\begin{tabular}{cp{6cm}}
\hline

$n$& The index of user. \\
\hline
$m$& The index of slice.\\
\hline
$k$& The index of slicing window.\\
\hline
$Q_{m,k}$& The SLA satisfaction ratio of slice $m$ over slicing window $k$.\\
\hline
$o_{m,k}$& The isolation degree of slice $m$ over slicing window $k$.\\
\hline
$v_{m,k}$& The resource utilization ratio of slice $m$ over slicing window $k$.\\
\hline
$w_{m,k}$& The resource allocated to slice $m$ at slicing window $k$.\\
\hline
$w_{m,c,k}$&The resource that slice $m$ occupies from the common slice $w_{c,k}$.\\
\hline
$S_k$& The spectrum efficiency of slicing window $k$.\\
\hline
$p_{n,k}$ & the position of user $n$ at the end of slicing window $k$. \\
\hline
$d_{n,k}$ & The total traffic demand of user $n$ at the end of slicing window $k$.\\
\hline
$a_{m,k}$ & The resource adjustment action of slice $m$ at the start of slicing window $k$.\\
\hline
\end{tabular}
\label{tab:symbol}
\end{table}

We consider a typical OFDMA based downlink cellular network consisting of multiple base stations (BSs), denoted by $\mathcal{B} = \{1,2,\cdots,B \}$, as shown in Fig. \ref{fig:sys}. BSs serve multiple users, which is denoted as the set $\mathcal{N} = \{1,2,\cdots,N \}$. Assume that the cellular network consists of a set of network slices denoted as $\mathcal{M} =  \{1,2,\cdots,M\}$ and each user equipment (UE) can associate with one or more slices of $\mathcal{M}$ and $\mathcal{N}_m$ denotes UEs that belong to slice $m$.
Radio resource is divided into Transmission Time Intervals (TTIs) denoted by $t\in \{1,2,\cdots\}$ in the time domain. {$W_b$ represents the number of resource blocks (RB) partitioned from the bandwidth of BS.} The duration of a slicing window, where the resource allocated to each slice remains constant, is called an \textit{slicing window}, denoted by $k\in\{1,2,\cdots\}$, and each {slicing window} contains $T$ consecutive TTIs. {Key notations used in this paper are presented in Table \ref{tab:symbol}.} Considering an equal power allocation, the received SINR of user $n$ associated with BS $b$ at time $t$ is given as
\begin{equation}
\gamma_{n,t} = \frac{P_b H_{b,n,t}}{W_bN_0},
\end{equation}
where $P_b$ is the transmit power of BS $b$ and $H_{b,n,t}$ is the channel gain between BS $b$ and user $n$. $N_0$ is the power of additive white Gaussian noise.

For the traditional traffic with a large packet size, e.g. eMBB traffic, the achievable rate of the user $n$ can be directly estimated according to Shannon's capacity {\cite{jie2021tcom,shannon1948mathematical}}. For the short-sized packet transmission, such as URLLC and MTC services, the data rate falls in the finite blocklength channel coding regime \cite{polyanskiy2010channel}. Therefore, the data rate of UEs is denoted as \eqref{datarate},
\begin{figure*}[t]
\begin{equation}
\label{datarate}
r_{n,t} = \left \{ \begin{array}{ll}
       W_{n,t}\log_2\left(1+\gamma_{n,t} \right), &\text{for long packets transmission}\\
       W_{n,t} \left[{\rm log} \left( 1+\gamma_{n,t} \right) - \sqrt{\frac{C_{n,t}}{l_{n,t}}}Q^{-1}\left( \epsilon \right) {\rm log} e\right], &\text{for short packets transmission}
\end{array}
\right.
\end{equation}
\end{figure*}
where $W_{n,t}$ is the allocated bandwidth to UE $n$ within $t$-th TTI. $\epsilon$ is the transmission error probability, and $Q^{-1}\left(\cdot \right)$ is the inverse of the Gaussian Q-function, and $l_{n,t}$ represents the length of codeword block in symbols, and $C_{n,t}$ is channel dispersion, given by $C_{n,t} = 1-\frac{1}{\left( 1+\gamma_{n,t}\right)^2}$.

\subsection{Slice and Mobility Model}
As shown in Fig. \ref{fig:sys}, we consider a mixed slices' traffic scenario consisting of traditional traffic, e.g., file and multimedia traffic, and C-V2X traffic in Mode-3, where BSs, e.g., eNodeB (eNB) and gNodeB (gNB), directly allocate radio resources to vehicles for their V2X communication through Uu interface (Uu interface refers to the link between UE to the terrestrial radio access network) in a centralized way. The former is typical Enhanced Mobile BroadBand (eMBB) traffic which is latency and reliability tolerant. Its SLA focuses on the minimum throughput. On the contrary, the SLA of C-V2X in terms of latency and reliability are very stringent compared with eMBB, with ultra-reliability and low latency, such as 99.999\% and 1 ms in \cite{3gpp.38.824}. In this work, we consider two slices: eMBB and C-V2X Ultra-Reliable Low-Latency Communication (URLLC), as shown in Fig. \ref{fig:sys}. The BS reserves a number of RBs for each slice based on their SLA requirements and objectives: high throughput for eMBB, low latency and high packet reliability for URLLC, a certain isolation degree for each slice.

An overview of the mobility model of UEs is given in Fig. \ref{fig:sys}. For the eMBB UEs, we consider a low-speed mobility model, such as the pedestrian of people. However, the motion of vehicle UEs is faster than eMBB UEs. The vehicle is driven along the road from one place to another. Their motion is affected by intersections, traffic lights and routes. Denote the position of UE $n$ at the end of {slicing window} k as $p_{n,k}$.

\subsection{SLA Model}

Generally speaking, classical QoS metrics for slices' SLA include throughput, packet latency and transmission reliability. For the throughput, it can be easily derived by aggregating the amount of data that is successfully transmitted over time. For the packet delay, a detailed queuing model of UEs' packets needs to be clarified.

In this paper, the arrival distribution of traffic is characterised by the pattern of service, and there is no prior knowledge of volatile demand. The arriving packets of UEs are cached in the BS's buffer and are delivered according to the first-come-first-serve (FCFS) policy. Assume that each UE is corresponding to one data queue at BS. The packet delay consists of two parts, i.e., queuing time and transmission time, where the former is influenced by scheduling policy and the latter is decided by instantaneous data rate. For example, the delay of the $i$-th packet of UE $n$ is calculated as
\begin{equation}
    D_{n,i} = W_{n,i}+\delta_{n,i},
\end{equation}
where $W_{n,i}$ is the queuing time of the $i$-th packet and $\delta_{n,i}$ is the transmission time. For the slice $m$, the queuing time is close to zero if the average packet arrival rate of UEs is low. And the packet delay is mainly decided by transmission time. With the increase of packet arrival rate, the packet delay is determined by both the queuing and transmission time.

From the perspective of the network, the packet is dropped if its delay exceeds the predefined maximum packet latency \cite{netw2020mei}. The reliability is determined by the percentage of packets that are successfully delivered. Therefore, the transmission reliability of UE $n$ is expressed as
\begin{equation}
    \theta_{n} = \text{Pr}\{ D_{n,i} \leq D_m^{max} \}, n\in\mathcal{N}_m, 
\end{equation}
where $D_{n,i}$ is the delay of the $i$-th packet of UE $n$, and $D_m^{max}$ corresponds to the maximum packet delay of UEs in slice $m$. 

{For the eMBB slice, its main performance metric is the throughput. In this paper, its SLA satisfaction ratio is defined as the ratio of the achieved data rate to the required rate specified in the SLA. For the C-V2X slice, its performance metrics are latency and reliability. The corresponding SLA satisfaction ratio is defined as the percentage of packets that are reliably transmitted within the specified delay in the SLA. Therefore, the SLA satisfaction ratio of slice $m$ within one slicing window $k$ is denoted as \eqref{SLA_q}, where $R_{m}^{th}$ is the minimum data rate requirement of slice $m$. $\theta_{n}^k$ represents the transmission reliability of packets of UE $n$ within slicing window $k$ and $D_m^{max}$ denotes the specified maximum packet delay.} 

Thus, we use the throughput, latency and reliability as the QoS metrics to evaluate the SLA satisfaction in the following.

\begin{figure*}[t]
\begin{equation}
\label{SLA_q}
Q_{m,k} = \left \{ \begin{array}{ll}
       \frac{1}{\left|\mathcal{N}_m \right|} \sum_{n\in \mathcal{N}_m} \text{min}\left( \frac{\sum_{t=(k-1)T+1}^{kT}r_{n,t}}{R_{m}^{th}},1\right), &\text{for the eMBB slice}\\
       \frac{1}{\left|\mathcal{N}_m \right|}\sum_{n\in \mathcal{N}_m} \theta_{n}^k, &\text{for the uRLLC slice}
\end{array}
\right.
\end{equation}
\end{figure*}

\subsection{Impact of High Mobility on Slicing}

As mentioned before, the resource dedicated to one slice remains constant during a slicing window. Depending on the dynamic of the environment and realization of network slicing, the slicing window can be configured with different time granularity, e.g., milliseconds, seconds, minutes and hours \cite{ISRRA2020WC}. Generally speaking, the longer slicing window represents a higher level of isolation and lower resource flexibility, while the shorter implies the opposite.

In a high mobility scenario, when a UE moves to a new cell, if the resources corresponding to its slice are sufficient, it can be instantaneously served as enabled by the seamless handover enhancements of 5G {new radio} (NR) \cite{3gpp.38.913}. However, if the resources of the slice are not sufficient, the SLA of the corresponding slice deteriorates significantly. Intuitive ideas are configuring the slicing window as short as possible or implementing resource over-provision. However, both methods have their challenges. On the one hand, too short slicing windows bring frequent resource slicing reconfiguration, which in turn increases network complexity. Furthermore, limited hardware capacity and practical signalling process do not allow for slicing window configuration in a too short time level, e.g., millisecond level. For example, such reconfiguration involves {radio resource control }(RRC) procedures which introduce about 80-100 ms delay \cite{3gpp.36.331}. On the other hand, the over-provision method results in low SE.

To guarantee the SLA requirement of high mobility scenarios and improve the SE, we first propose a hard and soft hybrid slicing framework in Section \ref{sec:proformu}. Then, a two-step DL-based RAN resource slicing is given in Section \ref{solution}.

\section{Hybrid Slicing Framework and Problem Formulation }
\label{sec:proformu}
In this section, we first propose a hard and soft hybrid slicing framework. Then, we formulate the IS-RRA problem as a utility maximization problem.
\begin{figure}[t]
    \centering
    \includegraphics[width=0.45\textwidth]{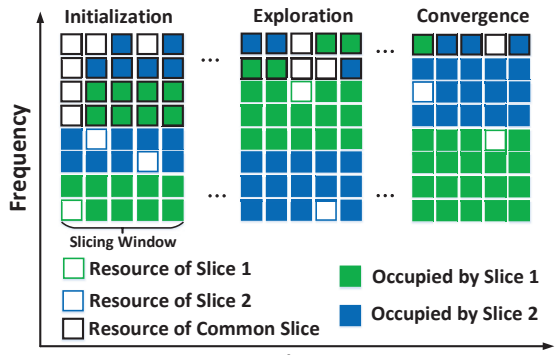}
    \caption{The illustration of the hybrid slicing framework.}
    \label{fig:hybrid_framework}
\end{figure}

\subsection{Hybrid Slicing Framework }
The purely hard slicing strategy, where each slice can only occupy the resource allocated to it, can guarantee full isolation among slices while it results in low SE. On the contrary, the soft slicing method driven by flexible resource sharing can maximize SE while limited by isolation. Therefore, we propose a novel hybrid slicing framework that can take advantage of both hard and soft strategies. Especially, soft decision, i.e. common slice setting, is proposed to guarantee SLA and improve resource efficiency in high mobility scenarios and the exploration phase of Section \ref{solution}. The hybrid slicing framework can be understood from the following two aspects.
 
 {\em 1) Common Slice Setting}: 
Fig. \ref{fig:hybrid_framework} shows a hybrid scheme, where the resources are divided into two parts, i.e. resources dedicated to slices and resources of the common slice, corresponding to hard and soft strategies. For the hard part, the resource allocated to each slice can only be occupied by UEs of the corresponding slice. For the soft part, all UEs can utilize the resource of the common slice according to their demand and priority.

 In a high mobility scenario, the necessary RBs in worst-case situations, e.g., when UEs move to the cell edge or new UEs move to the cell, are more than the average occupied RBs over the entire slicing window. The general solution of hard strategy is to allocate the required RBs of the worst-case to guarantee the SLA of the entire slicing window, which leads to a low SE. However, the reasonable resource configuration of the hybrid scheme enables both SLA satisfaction and resource efficiency with a small sacrifice of isolation. For example, $80\%$ resources required in worst-case are allocated to each slice as the hard part, which can realize SLA guarantee in most TTIs of slicing window. Moreover, each slice can occupy the RBs of the common slice when the hard part cannot satisfy its QoS metric. Therefore, the optimal resource configuration of the hybrid scheme, e.g., the right part of Fig. \ref{fig:hybrid_framework}, can achieve the SLA guarantee and maximize the SE under a specific isolation constraint.

{\em 2) Periodically Adjusting Resource Slicing}: {As Fig. \ref{fig:hybrid_framework} shows, the network allocates resources for each slice at the beginning of each slicing window. Therefore, radio resources among slices can be periodically adjusted to adopt a dynamic wireless environment. Generally speaking, the hybrid slicing framework will go through three stages, namely \textit{initialization}, \textit{exploration} and \textit{convergence}, as shown in Fig. \ref{fig:hybrid_framework}. In the initialization phase, due to the unknown wireless environment, the soft decision part of the network will dominate to guarantee the slices' SLA and cannot achieve the required performance isolation. Then, the network enters the exploration phase and continuously adjusts the inter-slice resource allocation to approach the SLA and isolation requirements based on the available performance feedback. As the exploration phase progresses, the network's awareness of the wireless environment gradually increases and eventually, the slice resource allocation enters the convergence phase, where slices' SLA and isolation can be satisfied. }
    
\subsection{Problem Formulation}
\label{proformu}
For a slice $m$, the degree of isolation in {slicing window} $k$ is represented by follows
\begin{equation}
    o_{m,k} =  \frac{w_{m,k}}{w_{m,k}+w_{m,c,k}}
\end{equation}
where $w_{m,k}$ is the allocated resources of slice $m$ and $w_{c,m,k}$ denotes resources that slice $m$ occupies from the common slice $w_{c,k}$. The objective of the RAN slicing is to guarantee the SLA of diverse slices and simultaneously maximize the SE, which are defined as follows
\begin{eqnarray}
Q_{m,k} &=& f\left(d_{m,k},{w_{m,k},w_{m,c,k}} \right), \label{Q_mk}\\
S_k &=& \sum_{t=\left(k-1\right)T+1}^{kT}\sum_{n=1}^{N}\frac{r_{n,t}}{W} 
\label{S_k}
\end{eqnarray}
 where $d_{m,k}$ is the fluctuation traffic demand of slice $m$, and function $f\left(\cdot\right)$ represents the complicated relationship between the SLA and traffic demand, allocated resources to slices and scheduling algorithms within slices.

 The utility function of one {slicing window} is defined as follows
\begin{equation}
\label{utility}
    U^k =  \sum_{m=1}^{M}\alpha_m Q_{m,k} +\beta\prod\limits_{m=1}^{M}\mathbbm{1}(Q_{m,k})\cdot S_{k},
\end{equation}
where $\alpha_m$ is the utility coefficient of slice $m$ and $\beta$ corresponds to SE. $\mathbbm{1}(Q_{m,t})$ is the indicator function to denote whether the SLA of slice $m$ is satisfied. Define the threshold of SLA of slice $m$ as $Q_{m}^{th}$, we have
\begin{eqnarray}
\label{Qindi}
\mathbbm{1}(Q_{m,k}) = \left \{ \begin{array}{ll}
      1, &\text{if } Q_{m,k}\geq Q_{m}^{th} \\
      0, &\text{otherwise}
\end{array}
\right.
\end{eqnarray}

The objective of a slice network is to maximize the long-term utility. A general method to maximize the average utility within a finite time period $K$, e.g., an hour, a day, or a week \cite{8931583}. Hence, the network slice problem is formulated as follows.
 \begin{eqnarray}
\mathcal{P}: \underset{w_{m,k},w_{c,k}}{\textrm{max}} && \frac{1}{K}\sum\limits_{k=1}^{K} U_k \\
\textrm{s. t.} && \eqref{Q_mk},\eqref{S_k} \nonumber \\
&& o_{m,k} \geq o^{th}_{m}, \label{omega}\\
&& \sum_{m \in \mathcal{M}} \ w_{m,k}+w_{c,k} = W  \label{equa:w_ms},
\end{eqnarray}
where $o^{th}_{m}$ represents the threshold of required isolation. 

The difficulties of problem $\mathcal{P}$ is reflected in the following aspects.
\begin{itemize}
    {\item \textit{Heterogeneous QoS}: The heterogeneous QoS, i.e., throughput, packet delay, reliability, of slices highly complicates the problem.} 
    
    {\item\textit{Highly dynamic environment}: Due to the varying network dynamics caused by high mobility, it is very tricky to satisfy the strict latency and ultra-reliability requirements of C-V2X UEs. Accordingly, such a situation makes it intractable to guarantee the SLA of the C-V2X slice.}
    
    {\item\textit{Customized scheduling algorithms}: Each slice can adapt the customized scheduling algorithm within the slice according to specific QoS metric. The customized scheduling algorithms and volatile traffic demand make $f\left(\cdot\right)$ extremely complex. An analytical model of $f\left(\cdot\right)$ in practical networks is almost impossible to derive \cite{mao2019learning}.}
    
    {\item \textit{Markovian characteristics of resource slicing}: The inter-slice resource allocation of slicing systems exhibits \textit{Markovian} characteristics, i.e. the allocation strategy affects not only the current SLAs and resource efficiency but also further network state and utility, e.g., the queue of UEs and delay of packets.}
\end{itemize}
To address the challenges above, we proposed a prediction-based IS-RRA through a two-step DL solution in the next Section.

\section{Deep Learning-based Proactive Solution}
\label{solution}
\begin{figure}[t]
    \centering
    \includegraphics[width=3.2in]{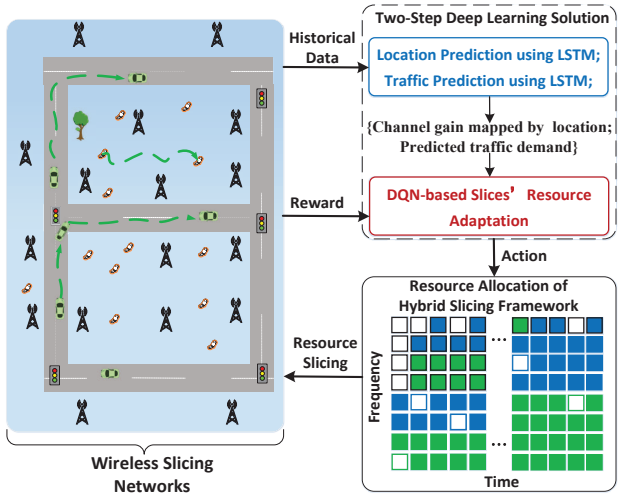}
    \caption{{The overview of the proposed two-step deep learning-based solution.}}
    \label{fig:solu}
\end{figure}

A prediction-based slicing framework through a two-step DL scheme is proposed to solve the above problem (\textbf{Problem} $\mathcal{P}$). As shown in Fig. \ref{fig:solu}, both the LSTM neural network and DRL are incorporated in the two-step DL solution.

First, LSTM networks are adopted to predict the network state, including location and traffic of UEs in {slicing window} level according to historical data. Then we map the location to channel gain by a perfect radio map that records pathloss and shadowing at different locations \cite{outdoorsurvey}. It is worth noting that the predicted channel gain and traffic demand are on {slicing window} level corresponding to a large-time scale. Therefore, a precise scheme cannot be directly available by the predicted channel gain and traffic demand. Considering the Markovian characteristic of the problem $\mathcal{P}$, as well as the superiority of DRL in model-free problems, {a DQN-based solution is proposed to give the final resource allocation of different slices based on the proposed hybrid slicing framework. Especially,} the predicted channel gain and traffic demand of next {slicing window} and real channel gain and traffic of current {slicing window} are jointly input to the DQN as the state to output the resource slicing of next {slicing window}. Finally, the DQN learns a slicing policy to achieve optimal utility.

\subsection{LSTM Based Location and Traffic Prediction} 

Before the start of each {slicing window}, the location and traffic demand of each UE is predicted to promote a seamless resource adjustment in a high mobility scenario. To predict the location and traffic demand, a central processor is deployed in the cloud that stores the historical data, predicts the location and traffic of each UE before the start of each {slicing window}.

The historical data of location and traffic of user $n$ is denoted as $\mathbf{p}_{n,k}^{his} = \left[p_{n,k-Z+1},\cdots,p_{n,k-1}\right]^T$ and $\mathbf{d}_{n,k}^{his} = \left[d_{n,k-Z+1},\cdots,d_{n,k-1}\right]^T$, where $d_{n,k}$ represents the total traffic demand of user $n$ over {slicing window} $k$ and $Z$ is the number of samples in the history. Consider a one-step prediction, the prediction results is denoted as $p_{n,k}^{pre}$ and $d_{n,k}^{pre}$, respectively.

Considering the effectiveness of LSTM on forecasting the continuous-time series, e.g., traffic prediction in \cite{trafficpre2020icc} and user mobility prediction in \cite{locapre2021TITS,vehiclepre2021TITS}, we apply an LSTM-based neural network to predict traffic and location for each user with historical data records $\mathbf{p}_{n,k}^{his}$ and $\mathbf{d}_{n,k}^{his}$. 

Two LSTM networks with the same architecture referring to \cite{trajecoty2018ICSP} are designed and trained for traffic and mobility prediction, respectively. The input of the LSTM is denoted as $\mathbf{x}_{n,k}^{his}$, i.e., $\mathbf{p}_{n,k}^{his}$ or $\mathbf{d}_{n,k}^{his}$, and the output is $x^{pre}_{n,k}$, i.e., $p_{n,k}^{pre}$ and $d_{n,k}^{pre}$, respectively. The relation between input and output can be written as 
\begin{equation}
   x^{pre}_{n,k} = g\left(\mathbf{x}^{his}_{n,k}; \psi\right) 
\end{equation}
where $g\left(\cdot\right)$ and $\psi$ represent the nonlinear transformation and the parameters of LSTM networks. With labeled historical data $\{\mathbf{x}_{n,k}^{his}, x_{n,k}\}$ {where $x_{n,k}$ denotes the ground truth of user $n$ in {slicing window} $k$, i.e., $p_{n,k}$ or $d_{n,k}$}, LSTM networks {are} trained by minimizing the mean-square-error (MSE) loss as follows,
\begin{equation}
    \epsilon\left(\psi\right) = \frac{1}{NK}\sum_n\sum_k \left(x_{n,k}-x_{n,k}^{pre}\right)^2.
\end{equation}

After LSTM networks are trained well, the online prediction outputs the location and traffic of the next {slicing window}. Furthermore, the channel gain of each user can be mapped through location and a radio map. Then the predicted channel gain $h_{n,k}^{pre}$ and traffic $d_{n,k}^{pre}$ are sent to the DRL network to derive the final resource slicing.

\subsection {Deep Reinforcement Learning-based Slicing} 
As mentioned before, the predicted channel gain and traffic are utilized by DRL to give the {refined} policy to achieve the optimal utility over the entire {slicing window}. In this paper, an initial resource allocation of slices, e.g., combining NVS \cite{kokku2011nvs} and the predicted channel gain and traffic information, is first given. Then the DRL agent dynamically adjusts the resource allocated to slices to guarantee the SLA and isolation of slices. To achieve efficient and intelligent slicing, the agent observes the environment, i.e., the dynamics of channel gain and traffic of contiguous {slicing windows}, and makes a decision according to the observed state at the start of each {slicing window}. The states, actions and reward of the DRL scheme are defined as follows. 

\textbf{State:} The state at {slicing window} $k$ contains the channel and traffic information of last {slicing window} $k-1$ and the predicted ones of $k_{th}$ {slicing window}, and upper layer performance feedback information of {slicing window} $k-1$. The DRL network is expected to learn the resource adjustment strategy by observing the channel and traffic variations of two adjacent {slicing windows} and the performance feedback of the last {slicing window}.

Considering the dynamic distribution caused by user mobility, directly using the channel and traffic information of the users as the input to the network will lead to dynamic input dimensions, making the network untraceable. Therefore, we introduce a hierarchical channel gain set $\{c_1, \cdots, c_j, \cdots,c_J\}$ where $c_1<\cdots<c_J$. The $h_{m,k,j}$ denote the number of users in slice $m$ at {slicing window} $k$ whose channel gain is between $c_j$ and $c_{j+1}$. Similarly, $d_{m,k,j}$ denotes the total traffic demand of the corresponding users. Based on these factors, the state is defined as a tuple as follows
\small
\begin{equation}
  s_{k} = \{\mathbf{h}_{m,k}, \mathbf{d}_{m,k},\mathbf{h}_{m,k}^{pre}, \mathbf{d}_{m,k}^{pre}, Q_{m,k}, o_{m,k}, v_{m,k}| m \in \mathcal{M}\} \nonumber
\end{equation}
\normalsize
where $\mathbf{h}_{m,k} = \left[h_{m,k,1},\cdots,h_{m,k,J}\right]$ and $\mathbf{d}_{m,k} = \left[d_{m,k,1},\cdots,d_{m,k,J}\right]$. $v_{m,k}$ denotes the resource utilization ratio of slice $m$ over the slicing window $k$.

\textbf{Action:} The agent intelligently adjusts the resource allocation of slices by selecting an action $a_k$ according to the current state $s_k$. The action for a slice is defined as decreasing, remaining or increasing the allocated resource. It is worth noting that the object of action interaction is the common slice. For example, slice $m$ offloads additional resources to the common slice, and slice $m+1$ require more dedicated resources from the common slice at {slicing window} $k$. The action set of one slice is defined as $\mathcal{A}=\{-a^i, \cdots,-a^1,0,a^1,\cdots,a^i\}$, where $0<a^1<\cdots<a^{i}$ and $i$ is the positive integer. For example, define the action of slice $m$ is $a_{m,k}$, where $a_{m,k} \in \mathcal{A}$, we have $w_{m,k+1}=w_{m,k}+a_{m,k}$. Therefore, the action of the agent at $k$ is the combination of actions of all slices, and it is defined as follows
\begin{equation}
    a_k = \{ a_{m,k} | m\in\mathcal{M}, a_{m,k} \in \mathcal{A}\}. \nonumber
\end{equation}

\textbf{Reward:} 
The reward of agent is defined as follows
\small
\begin{eqnarray}
r_k(s_k, a_k)&=&\alpha_m \sum_{m=1}^{M}e^{Q_{m,k}}+   \beta\prod\limits_{m=1}^{M}\mathbbm{1}(Q_{m,k})\cdot \frac{S_k}{S_{max}} \nonumber \\ 
&& -\rho\sum_{m=1}^{M}\left[o_{m}^{th} - o_{m,k} \right]^+ \nonumber ,
\end{eqnarray}
\normalsize
where $\left[x\right]^+ =\text{max}\left(0,x\right)$, and $\rho>0$ is a punishment constant. $\frac{S_k}{S_{max}}$ operation normalizes SE by dividing the predefined maximum value $S_{max}$. {The positive part of the reward is designed according to the utility defined in Section \ref{proformu}. Moreover, we incorporate the isolation constraint \eqref{omega} into the objective functions with the reward shaping technique \cite{griffith2013policy}. Therefore, there will be a penalty added to the reward if the isolation constraint is violated. Besides, the required SLA satisfaction ratio of the C-V2X slice can be very high, e.g., 99.99\%. Thus, by increasing the SLA satisfaction ratio of the C-V2X slice from 99\% to 99.99\%, the reward only increases 0.99\% if with a linear reward. As a result, the gradient of DQN in the training phases will be very small, which results in a long convergence time of DQN.} Therefore, the exponential reward function is {utilized} to train the network more efficiently as $Q_{m,k}$ approaches 1.

A DQN is applied to design and train the agent, where a neural network (NN) is used to approximate the action-value function, $q\left(s,a;\mathbf{\theta}\right) \approx Q^*\left(s,a\right)$ and $\theta$ represents the parameters of NN. The state is input to the DQN, and the network outputs the predicted Q values of each action. With the experience replay and quasistatic target network, the DQN is trained by minimizing the error between the predicted Q values and true Q values as follows,
\begin{equation}
L\left(\theta\right) = \frac{1}{B}\sum_k\left(y_k -q\left(s_k,a_k; \theta \right)\right)^2,
\end{equation}
where $B$ is the batch size. The target value $y_k$ is
\begin{equation}
y_k = r_{k+1} + \gamma\underset{a^{'}}{\text{max}}q\left(s_{k+1},a^{'};\theta^{'}\right),
\end{equation}
where $\theta^{'}$ represents the parameters of the target network and $\gamma$ is the discount factor. 

\begin{table}[b]
\centering
\caption{Slices Parameters}
\begin{tabular}{ccc}
\hline
\small
& eMBB& URLLC \\
\hline
Traffic Model & OAI data & period process\\
Packet Size & 6k bits & 256 bits\\
Arrival Rate  & 100 packets/s & 100  packets/s\\
SLA & 95\%\{5M bps\} & \{5ms, 99.99\%\} \\
$\alpha_m$ & 2 & 3\\
$o_m^{th}$ & 90\% & 90\% \\
Ave Num of UEs & 20 & 50\\
\hline
\end{tabular}
\label{tab:slice}
\end{table}

\section{Numerical Results}
\label{sec:experi}
\subsection{Experiments Setup}

We consider a road topology consisting of streets, intersections, traffic lights and routes, as shown in Fig. \ref{fig:sys}. There are nine road segments indexed by A$\thicksim$I and six intersections. The length of road A, \{C, D, F, G\}, \{B, E, H\}, I is 0.5 km, 1.5 km, 3 km, 1 km, respectively. The width of the road is 7 m. Consider a motion scenario: eMBB users randomly move with 1-2 m/s speed and V2X vehicles move from the point "Start" to the point "End" by randomly selecting three routes: \{A-B-D-G-I\}, \{A-C-E-G-I\} and \{A-C-F-H-I\}. The speed of vehicles ranges from 40 km/h to 70 km/h, and the acceleration and deceleration are 2 m/$s^2$ and 4 m/$s^2$, respectively. Vehicles ignore the light when they turn right, while they randomly stop 1$\thicksim$60 s when they go straight and turn left with a red light. The speed is 20 km/h when vehicles cross the intersection. The arrival rate of vehicles at point "Start" is 1 vehicle/s. 

The transmission power of each BS is $43$dBm, and interference among BSs is not considered in this work. $100$ RBs with each bandwidth $180$kHz are considered as total bandwidth resources for a BS. The pathloss model is consistent with our previous work \cite{myWCL2021}. Two slices corresponding to two types of services, i.e. eMBB and C-V2X services, are considered in the simulation.{ Assume that the coverage of each BS is 500 m and the handover delay is omitted. To obtain the ms level delay, the earliest deadline first (EDF) \cite{kargahi2005non} is utilized as the scheduling algorithm within the C-V2X slice and is performed in each ms. The eMBB slice adopts the classical round-robin (RR) algorithm.} We select one BS around the intersection, i.e. \textit{BS2} in Fig. \ref{fig:sys}, to evaluate the proposed algorithm. And the detailed slice parameters in one BS are summarized in Table \ref{tab:slice} to simulate a near-full load scenario. The values of $\gamma$, $\beta$ and $\rho$ are $0.9$, $5$ and $15$, respectively. {Considering the more stringent SLA requirements for the C-V2X slice, we set its reward factor $\alpha$ to be larger than the eMBB slice, as shown in Table \ref{tab:slice}.} The network architectures refer to LSTM in \cite{trajecoty2018ICSP} and DQN in \cite{DLMA}. The hierarchical channel gain set is $\{0,2,4,6,8,10,12,14\}${dB}. The resource allocated to the common slice is 40 RBs in the initial phase, and the action set for one slice is $\{-5,-2,0,2,5\}${RB}.
Three baseline algorithms are compared in our experiments:
\begin{itemize}
\item{\bf Optimal a Priori (OP):} Given a priori knowledge of traffic and SINR distributions of UEs, the optimal resource slicing is derived by exhaustive search. 

\item{\bf Hard-LSTM-A2C}\cite{DRLLSTM}: In this algorithm, a purely hard slicing framework that incorporates the LSTM into A2C is proposed to track user mobility and improve the system utility.

\item{{\bf Hard-DQN}}\cite{sun2019mix}{: In this algorithm, a purely hard slicing framework is proposed. And a centralized DQN is utilized to refine slice ratio of the whole network. And then the global slice ratio is mapped to each BS according to users’ SINR and data rate request. To eliminate the effect of imperfect mapping algorithms, in this paper we deploy the Hard-DQN algorithm at the single BS level.}
\end{itemize}
{In order to clearly compare the performance of different algorithms, the rewards of all compared algorithms are calculated based on the rewards designed in this paper. }

\begin{figure}[t] 
\centering  
\subfloat[{Location Prediction Error of C-V2X UEs}]{
\label{location}
\begin{minipage}[t]{2.8in}
\centering
\includegraphics[width=2.8in]{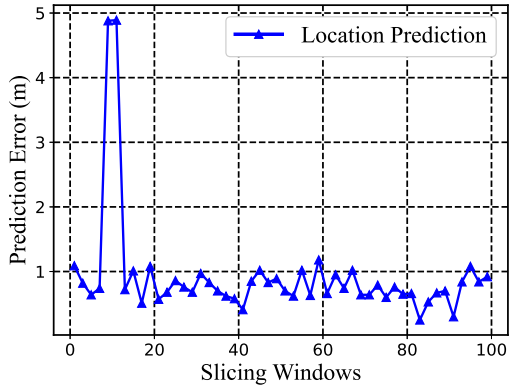}
\end{minipage}}

\subfloat[{Traffic Prediction of eMBB UEs}]{
\label{traffic}
\begin{minipage}[t]{2.8in}
\centering
\includegraphics[width=2.8in]{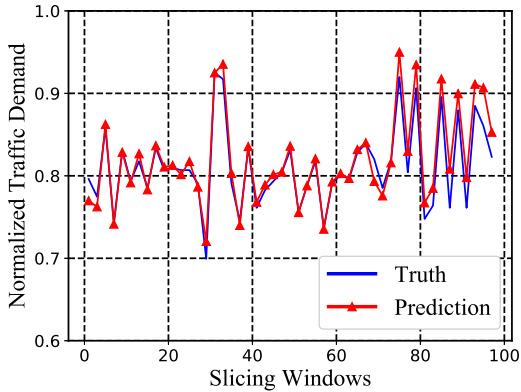}
\end{minipage}}
\caption{{The prediction results of location and traffic.}}
\centering
\label{fig:prediction}
\end{figure}

\subsection{Prediction Results}

The data set contains location and traffic historical data of two slices: 6000 trajectories of V2X UEs and corresponding length data of eMBB UEs. After LSTM networked is trained well, the online test results are shown in Fig. \ref{fig:prediction}. 

For the location prediction, Fig. \ref{location}. shows the results of one vehicle UE when it is within the coverage of the \textit{BS2} (see Fig. \ref{fig:sys}). We can see that the location prediction error is very small, i.e., around 1 m, except the $5_{th}$ and $6_{th}$ {slicing windows}. The data analysis reveals that on $5_{th}-15_{th}$ {slicing windows}, the vehicle is at the intersection. The direction randomness, e.g., the vehicle stops for the red light or turns right, and velocity variation increase the prediction error. However, the channel gain error caused by the 5 m location error can be negligible under a perfect radio map.

For the traffic prediction on an {slicing window} level, the prediction of periodic URLLC traffic is quite easy. Therefore, we give the prediction result of eMBB traffic in Fig. \ref{traffic}, where a real traffic dataset of high-dimensional live streaming generated by software-defined radio platform OpenAirInterface \cite{OAIweb}
is utilized. And the traffic dataset is used for eMBB UEs in all experiments. It can be observed that the error tends to increase when the traffic demand changes dramatically, e.g, the error from $75_{th}$ {slicing window} to $100_{th}$ {slicing window}. On the whole, large-scale traffic predictions are very accurate, and the normalized average prediction error is 0.07.

\begin{figure}[tbp]
    \centering
    \includegraphics[width=3 in]{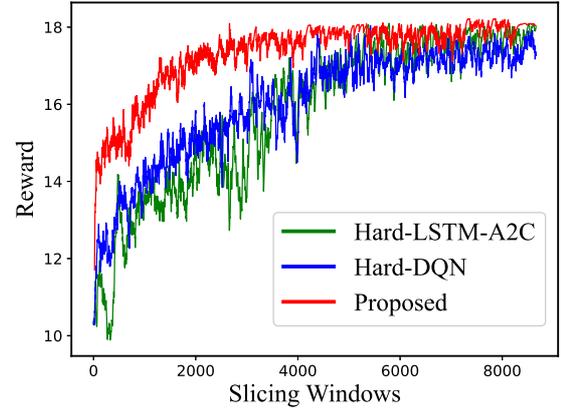}
    \caption{{The convergence process of two DRL based algorithms.}}
    \label{fig:train}
\end{figure}

\begin{figure}[!h] 
\centering  
\subfloat[{The proposed algorithm}]{
\label{iso1}
\begin{minipage}[t]{2.7in}
\centering
\includegraphics[width=2.7in]{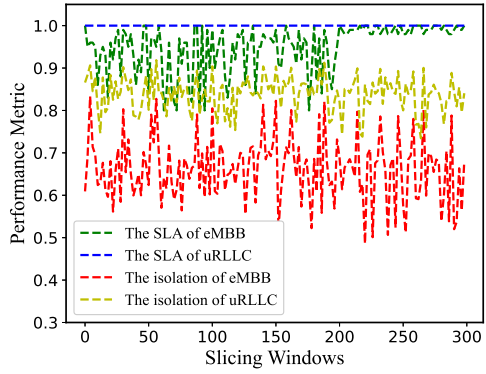}
\end{minipage}}

\subfloat[{Hard-LSTM-A2C}]{
\label{iso2}
\begin{minipage}[t]{2.7in}
\centering
\includegraphics[width=2.7in]{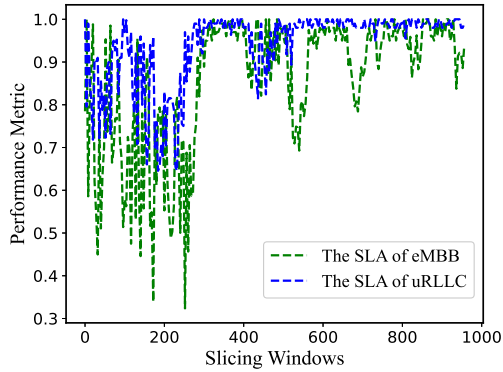}
\end{minipage}}

\subfloat[{Hard-DQN}]{
\label{iso3}
\begin{minipage}[t]{2.7in}
\centering
\includegraphics[width=2.7in]{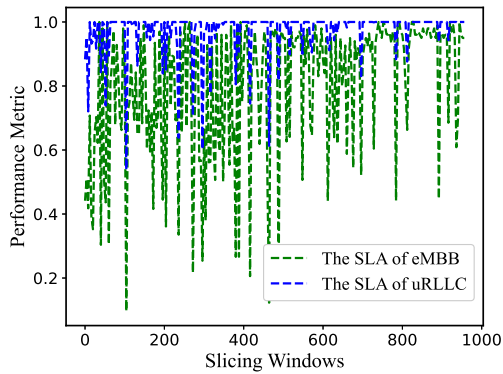}
\end{minipage}}

\caption{{The SLA satisfaction ratio of convergence process for the proposed DRL algorithm, the Hard-LSTM-A2C algorithm and Hard-DQN algorithm.}}
\centering
\label{fig:sla_iso}
\end{figure}

\subsection{The Analysis of DRL Convergence Process}
Fig. \ref{fig:train} illustrates the convergence process of { three} DRL-based algorithms in the experiments: { the proposed DRL algorithm, the Hard-LSTM-A2C algorithm and the Hard-DQN algorithm. First, the rewards of three DRL algorithms are low initially and increase with training until convergence.} Second, the superiority of the proposed algorithm over {the Hard-LSTM-A2C algorithm and the Hard-DQN algorithm} in the training process is shown in two aspects in the following. 

On the one hand, the reward of the proposed algorithm is significantly greater than that of the Hard-LSTM-A2C algorithm { and the Hard-DQN algorithm.} This is because DRL algorithms try some actions randomly to improve the policy in the exploration phase. Thus, the bad actions of random exploration of the Hard-LSTM-A2C {and the Hard-DQN algorithms} deteriorate the SLA satisfaction ratio significantly. However, the common slice setting and the limited action space of the proposed algorithm enable a much better SLA performance even in the exploration phase. 
For example, assume that the optimal resource allocation in the hard scheme is $\{75, 25\}$ RBs for the eMBB slice and C-V2X slice, respectively. In the exploration phase, the Hard-LSTM-A2C {and the Hard-DQN algorithms} may try some bad action like $\{10, 90\}$, $\{20, 80\}$ which dramatically deteriorates the slices' SLA and the reward. For the proposed algorithm, the initial action is based on a baseline algorithm and a bigger common slice setting, e.g., $\{50,10,40\}$ for the eMBB slice, C-V2X slice and common slice. Its exploration phase is a fine-tuning of the current allocation in each {slicing window}, such as increasing or decreasing the allocation of some slice. Thus, the bad action of it only slightly reduces the reward. Moreover, the bigger common slice setting of the initial phase greatly reduces the probability of extreme allocation, e,g., $\{10, 90, 0\}$. And the detailed SLA satisfaction ratio and isolation degree are shown in Fig. \ref{fig:sla_iso}.

On the other hand, the proposed DRL algorithm converges faster than the Hard-LSTM-A2C algorithm {and the Hard-DQN algorithms. Since the predicted channel gain and traffic demand serve as the part of state information. Moreover, the common slice setting increases the overall reward level. These two facts accelerate the convergence of the proposed algorithm.}

{Fig. \ref{fig:sla_iso} demonstrates the SLA satisfaction ratio of three DRL algorithms} in the early stage of the training process. Observing Fig. \ref{iso1}, the SLA of URLLC {of the proposed algorithm} is always guaranteed. The reasons lie in two aspects. First, the packet size of the URLLC slice is much smaller than the eMBB slice so that the required resource is lesser than the eMBB slice. Second, the shared resource of the common slice prevents extreme scenarios, e.g. most RBs are allocated to the eMBB slice. Similarly, the SLA of the eMBB slice is guaranteed after about 200 {slicing windows}. Compared with this, the URLLC slice's SLA of the Hard-LSTM-A2C algorithm can be guaranteed only after 500 {slicing windows}, {and the SLA satisfaction ratio of the URLLC slice of the Hard-DQN algorithm keeps fluctuating at the first 1000 {slicing windows}. For the eMBB slice, both two algorithms cannot achieve the required SLA satisfaction ratio at the first 1000 {slicing windows}. The reasons for the superiority of the proposed scheme over the other two compared algorithms have been elaborated in the discussion of Fig. \ref{fig:train}. That is the hybrid slicing setup prevents extremely harsh exploration actions and improves the SLA guarantee in the training phase.
In addition, the SLA satisfaction ratio in the Hard-DQN algorithm is lower and more volatile than that in the Hard-LSTM-A2C algorithm. This is because the latter utilizes the LSTM structure and can better track users' mobility.}

\begin{figure}[t]
    \centering
    \includegraphics[width=3in]{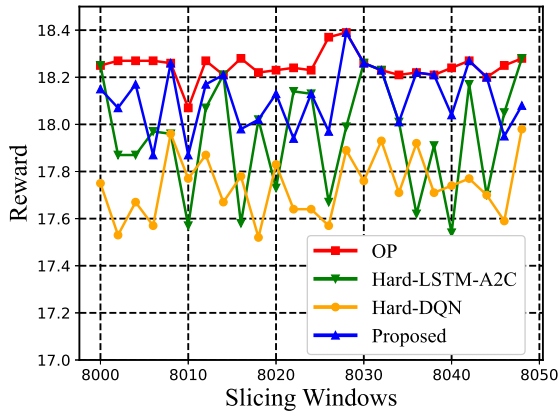}
    \caption{The rewards of four algorithms under 50 slicing windows after all algorithms are converged.}
    \label{fig:reward_com}
\end{figure}

We also give the isolation degree of the proposed DRL algorithm in Fig. \ref{iso1}. The isolation degree of {the Hard-LSTM-A2C and Hard-DQN algorithms is always 1, and we omit it in Fig. \ref{iso2} and Fig. \ref{iso3}.} Naturally, slices' isolation degree of the proposed DRL algorithm cannot approach the required thresholds, i.e., $90\%$, in the first 1000 {slicing windows}. However, it is pointless to discuss isolation when the slices' SLA cannot be guaranteed. Furthermore, the isolation degree of the proposed algorithm can achieve the required thresholds after the DRL network converges. 

\begin{table}[t]
\centering
\caption{The performance metric of four algorithms after all algorithms are converged.}
\begin{tabular}{|c|c|c|p{1.9cm}|c|}
\hline
\small
& \makecell[c]{OP}&\makecell[c]{Proposed}& \makecell[c]{Hard-\\ LSTM-A2C}& \makecell[c]{Hard-\\ DQN} \\
\hline
$Q_e$& $100\%$& $99.8\%$& $98.2\%$ &  {$95.4\%$} \\
\hline
$Q_u$ & $100\%$ & $100\%$ &$100\%$ &  {$100\%$}\\
\hline
$\upsilon_e$ & $99.0\%$ & $99.7\%$ &$90.5\%$ &{$94.3\%$}\\
\hline
$\upsilon_u$ & $81.4\%$ & $70.6\%$ &$62.4\%$ &{$53.7\%$}\\
\hline
$\upsilon_c$ & $99.4\%$ & $100\%$ & / & {/}\\
\hline
$o_e$ & $96.5\%$ & $97.9\%$ & $100\%$ & {$100\%$}\\
\hline
$o_u$ & $100\%$ & $99.4\%$ & $100\%$  & {$100\%$}\\
\hline
$\frac{S}{S_{max}}$ & 0.972 & 0.934 &0.916 &{0.872}\\
\hline
\end{tabular}
\label{tab:res}
\end{table}
\subsection{Performance Comparison}
Fig. \ref{fig:reward_com} shows the achievable reward of four algorithms from $8000_{th}$ {slicing windows} to $8050_{th}$ {slicing windows}, where {three} DRL-based algorithms have converged. It can be observed that both {the proposed DRL algorithm can achieve approximately optimal performance. And the reward of the proposed DRL algorithm is slightly larger than the Hard-LSTM-A2C and Hard-DQN algorithms. Since the common slice setting brings a bigger SE. Second, the Hard-LSTM-A2C algorithm outperforms the Hard-DQN algorithm a little. Since the LSTM structure utilizes the historical state information.}

Table \ref{tab:res} gives the average performance {of four algorithms from $8000_{th}$ slicing windows to $8050_{th}$ slicing windows} in terms of SLA satisfaction ratio, resource utilization ratio, isolation degree and normalized SE. $Q_e$ and $Q_u$ represent the SLA satisfaction ratio of the eMBB slice and C-V2X slice, respectively. Similarly, $\upsilon_e$, $\upsilon_u$, and $\upsilon_c$ denote the resource utilization ratio of the eMBB slice, C-V2X slice and common slice. $o_e$ and $o_u$ represents the isolation degree of two slices. We can see that both the proposed DRL algorithm and Hard-LSTM-V2C achieve near-optimal performance in terms of SLA satisfaction ratio, isolation and SE. The proposed prediction-based hybrid slicing framework exhibits a better performance on SLA satisfaction and SE. 
\section{Conclusion}
\label{sect:conc}

In this paper, we first propose a soft and hard hybrid slicing framework. Based on this, we propose a joint LSTM-based network state prediction and DQN-based network slicing adjustment strategy to adapt the user mobility. In the proposed solution, LSTM networks are trained to predict the traffic demand and location of each user. Moreover, channel gain is mapped by location and a radio map. Then, the predicted channel gain and traffic demand are input to the DQN to output the precise slicing adjustment strategy. The numerical results verify the effectiveness of the proposed DL-based hybrid slicing framework: 1) The hybrid slicing framework can not only balance the trade-off between isolation and SE of network slices but also significantly reduce the SLA violation in the DRL exploration phase; 2)It achieves near-optimal performance in terms of SLA satisfaction, isolation degree and SE.

\section*{ACKNOWLEDGEMENT}
\label{ACKNOWLEDGEMENT}
This work was supported in part by the National Natural Science Foundation of China (NSFC) under Grant 61871262, 62071284, and 61901251, the National Key R\&D Program of China grants 2017YFE0121400 and 2019YFE0196600, the Innovation Program of Shanghai Municipal Science and Technology Commission grant 20JC1416400, Pudong New Area Science \& Technology Development Fund, Key-Area Research and Development Program of Guangdong Province grant 2020B0101130012, Foshan Science and Technology Innovation Team Project grant FS0AA-KJ919-4402-0060, and research funds from Shanghai Institute for Advanced Communication and Data Science (SICS).

\bibliographystyle{gbt7714-numerical}
\bibliography{myref}
\biographies
\begin{CCJNLbiography}{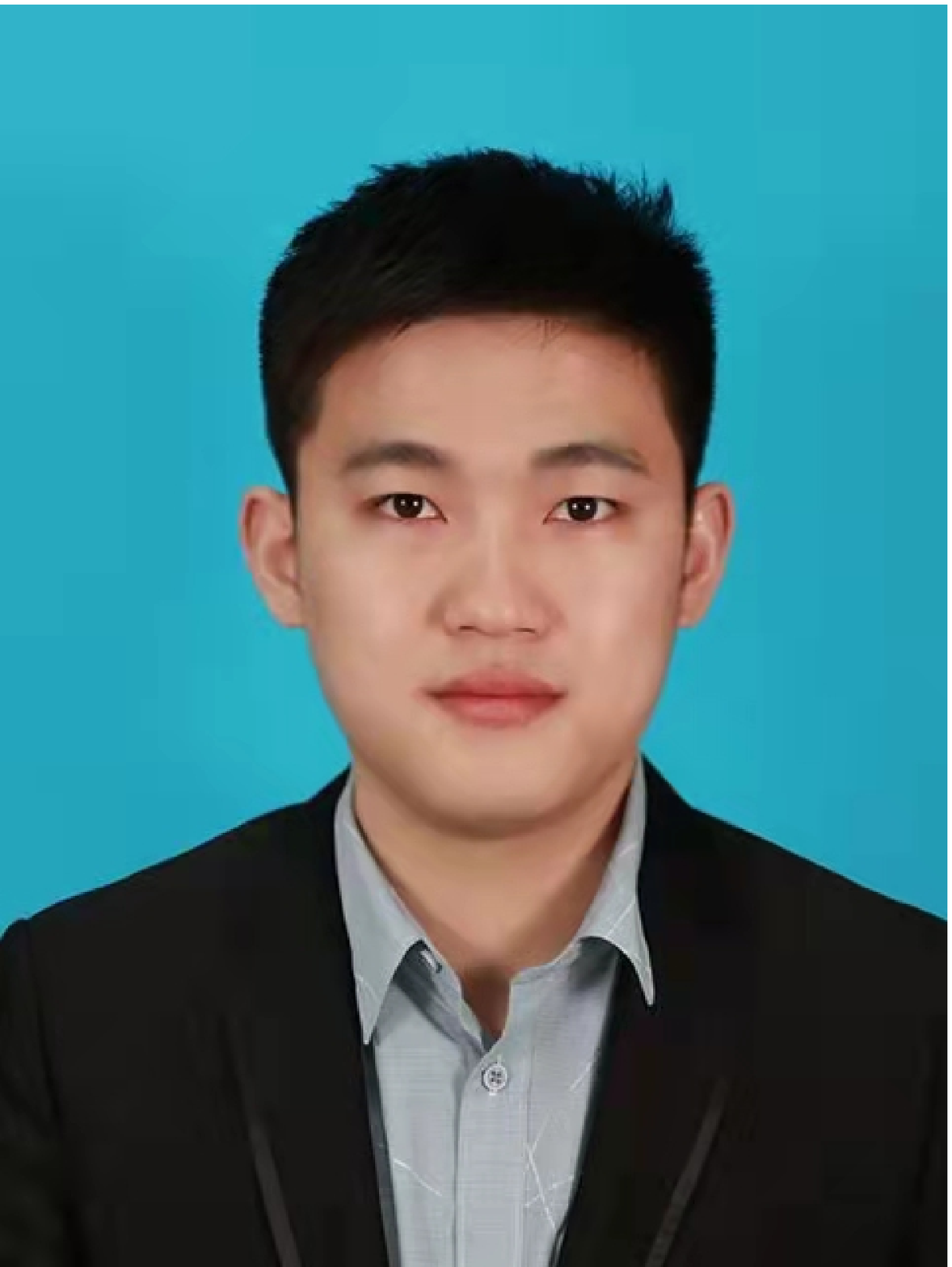}{Heng Zhang}
received the B.S. and Ph.D. degrees from the School of Communication
and Information Engineering of Shanghai
University (SHU) in 2017 and 2022, respectively. He is currently a Research Engineer with Huawei Technologies Company Ltd. His main research interests include wireless localization, network slicing and
resource allocation.
\end{CCJNLbiography}

\begin{CCJNLbiography}{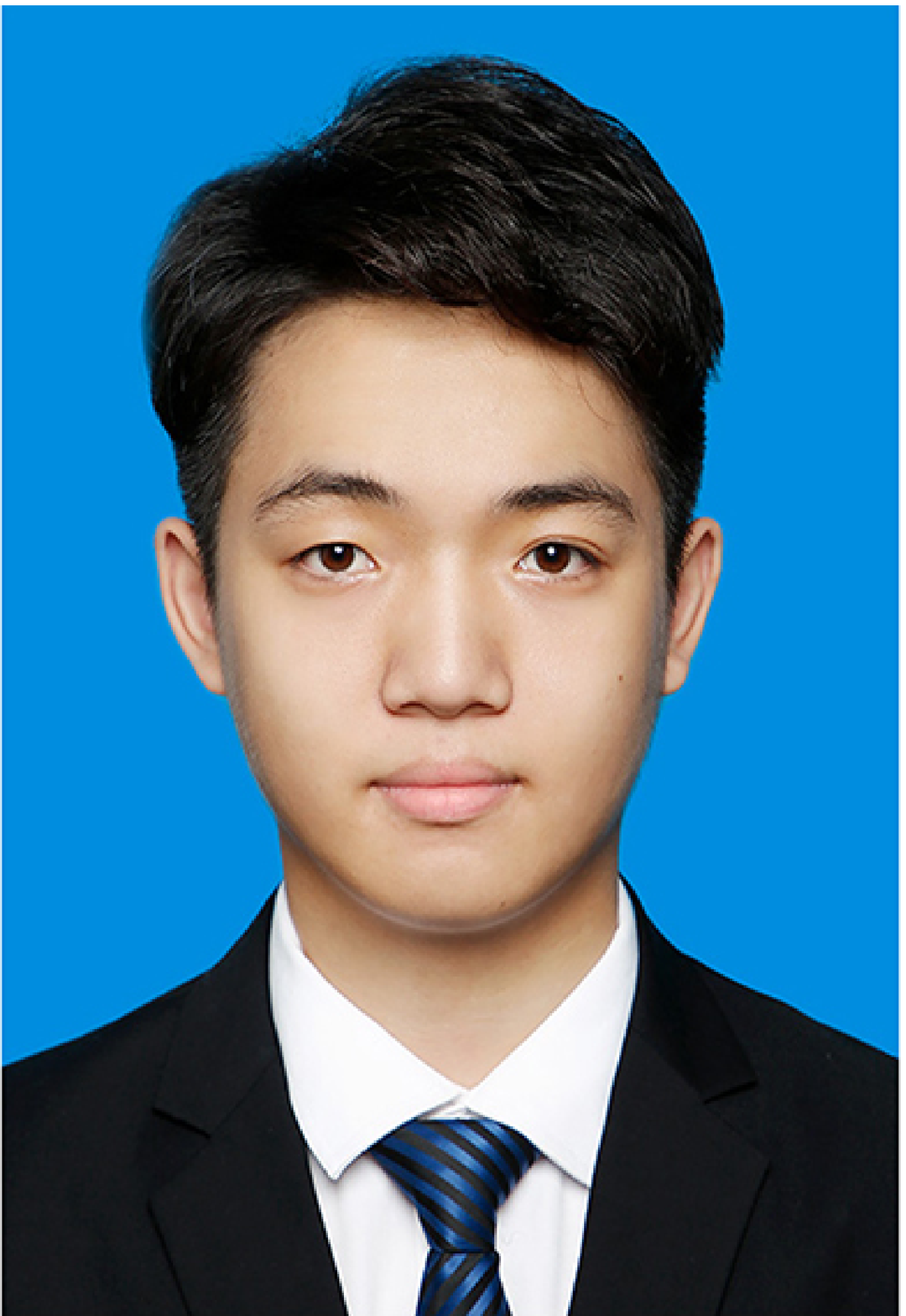}{Guangjin Pan}
received his B.E. degree from the School of Communication and Information Engineering of Shanghai University in 2018. He is currently studying for Ph.D. degree in the School of Communication and Information Engineering, Shanghai University. His main research interests include mobile edge computing, edge intelligence and wireless localization.
\end{CCJNLbiography}

\begin{CCJNLbiography}{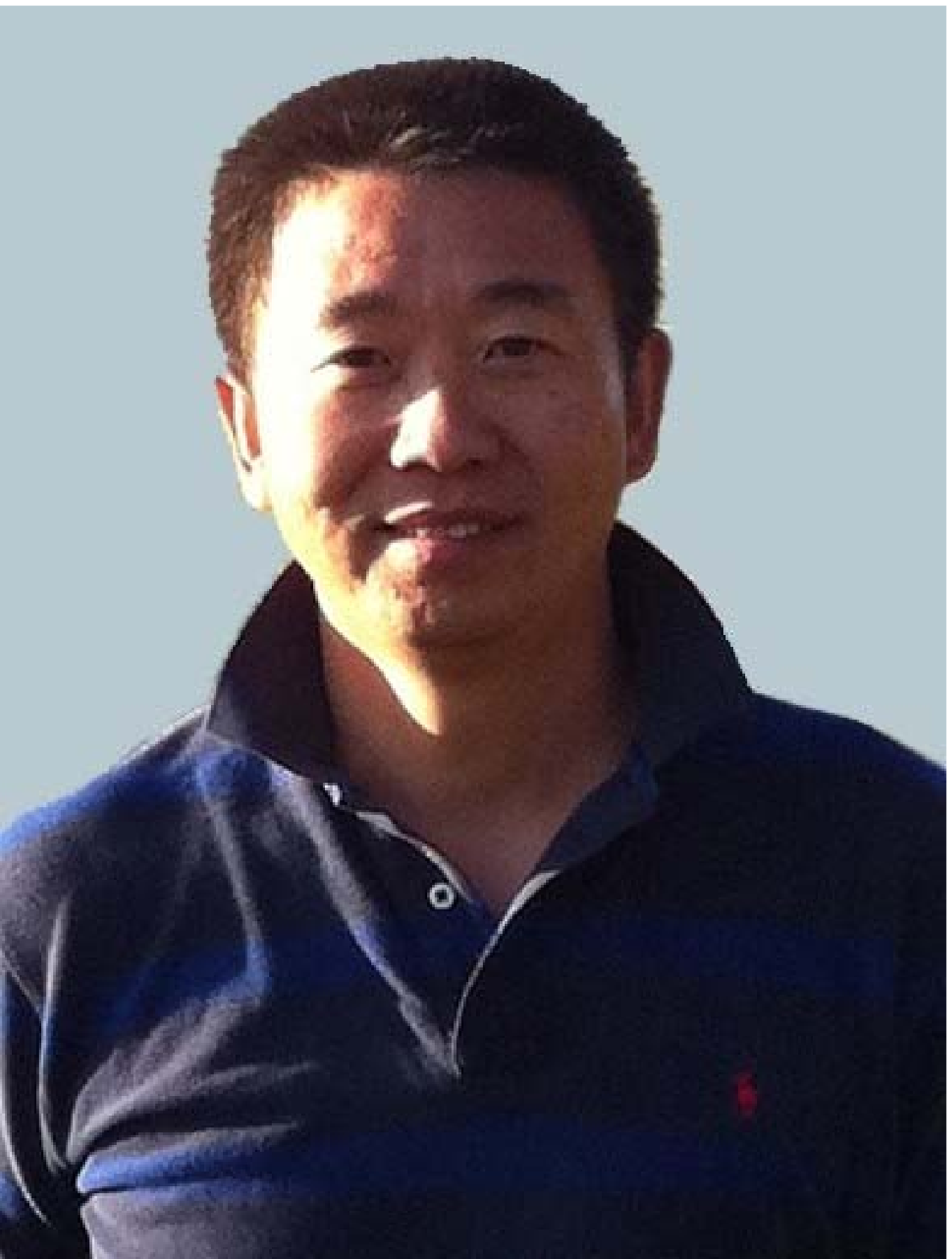}{Shugong Xu} (M’98-SM’06-F’16) graduated from Wuhan University, China, in 1990, and received his Master degree in Pattern Recognition and Intelligent Control from Huazhong University of Science and Technology (HUST), China, in 1993, and Ph.D. degree in EE from HUST in 1996. He is professor at Shanghai University, head of the Shanghai Institute for Advanced Communication and Data Science (SICS). He was the center Director and Intel Principal Investigator of the Intel Collaborative Research Institute for Mobile Networking and Computing (ICRI-MNC), prior to December 2016 when he joined Shanghai University. Before joining Intel in September 2013, he was a research director and principal scientist at the Communication Technologies Laboratory, Huawei Technologies. Among his responsibilities at Huawei, he founded and directed Huawei’s green radio research program, Green Radio Excellence in Architecture and Technologies (GREAT). He was also the Chief Scientist and PI for the China National 863 project on End-to-End Energy Efficient Networks. Shugong was one of the co-founders of the Green Touch consortium together with Bell Labs etc, and he served as the Co-Chair of the Technical Committee for three terms in this international consortium. Prior to joining Huawei in 2008, he was with Sharp Laboratories of America as a senior research scientist. Before that, he conducted research as research fellow in City College of New York, Michigan State University and Tsinghua University. Dr. Xu published over 100 peer-reviewed research papers in top international conferences and journals. He has over 20 U.S. patents granted. Some of these technologies have been adopted in international standards including the IEEE 802.11, 3GPP LTE, and DLNA. He was awarded ‘National Innovation Leadership Talent’ by China government in 2013, was elevated to IEEE Fellow in 2015 for contributions to the improvement of wireless networks efficiency. Shugong is also the winner of the 2017 Award for Advances in Communication from IEEE Communications Society. His current research interests include wireless communication systems and Machine Learning.
\end{CCJNLbiography}

\begin{CCJNLbiography}{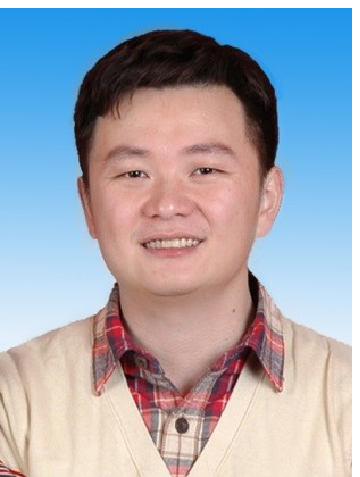}{Shunqing Zhang} (S’05-M’09-SM’14) received the B.S. degree from the Department of Microelectronics, Fudan University, Shanghai, China, in 2005, and the Ph.D. degree from the Department of Electrical and Computer Engineering, Hong Kong University of Science and Technology, Hong Kong, in 2009. He was with the Communication Technologies Laboratory, Huawei Technologies, as a Research Engineer and then a Senior Research Engineer from 2009 to 2014, and a Senior Research Scientist of Intel Collaborative Research Institute on Mobile Networking and Computing, Intel Labs from 2015 to 2017. Since 2017, he has been with the School of Communication and Information Engineering, Shanghai University, Shanghai, China, as a Full Professor. His current research interests include energy efficient 5G/5G+ communication networks, hybrid computing platform, and joint radio frequency and baseband design. He has published over 60 peer-reviewed journal and conference papers, as well as over 50 granted patents. He has received the National Young 1000-Talents Program and won the paper award for Advances in Communications from IEEE Communications Society in 2017.
\end{CCJNLbiography}

\begin{CCJNLbiography}{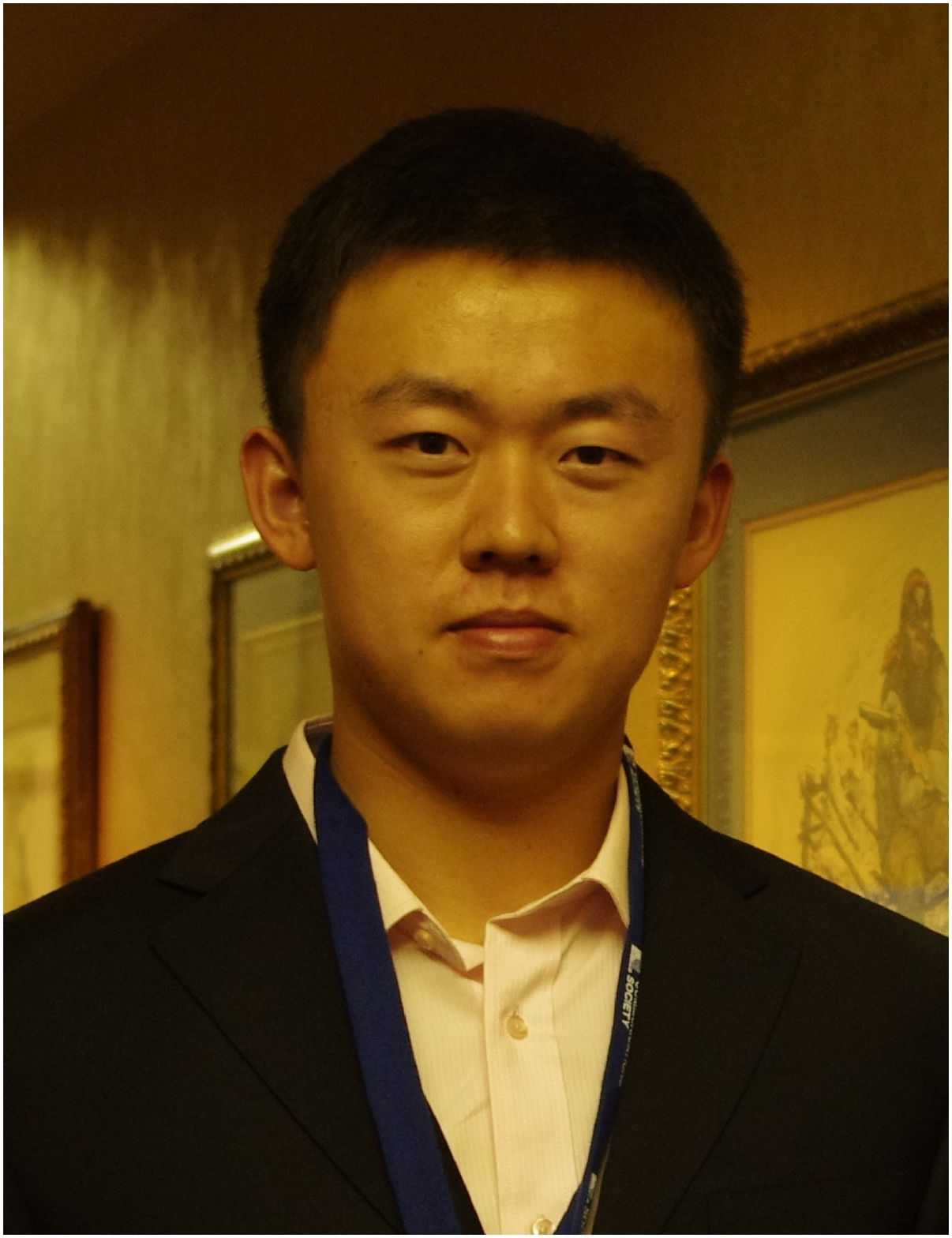}{Zhiyuan Jiang} (S’12-M’15) received his B.S. and Ph.D. degrees from the Electronic Engineering Department at Tsinghua University, China in 2010 and 2015, respectively. He is currently a Professor of School of Communication and Information Engineering in Shanghai University at Shanghai, China. He visited the WiDeS group at University of Southern California, USA from 2013 to 2014. He worked as an experienced researcher at Ericsson from 2015 to 2016. He visited ARNG at University of Southern California, USA from 2017 to 2018. He worked as a wireless signal processing scientist at Intel Labs, Hillsboro, USA during 2018. His current research interests include URLLC in wireless networked control systems and signal processing in MIMO systems. He serves as an associated editor for IEEE/KICS Journal of Communications and Networks, and a guest editor for IEEE IoT Journal. He serves as a TPC member of IEEE INFOCOM, ICC, GLOBECOM, WCNC and etc. He received the ITC Rising Scholar Award in 2020, the best paper award at IEEE ICC 2020, the best in-Session presentation award of IEEE INFOCOM 2019, and exemplary reviewer award of IEEE WCL in 2019.
\end{CCJNLbiography}

\end{document}